\documentclass[aip,rsi,reprint,graphicx]{revtex4-1} % for checking your page length
\usepackage{graphicx}
\usepackage{gensymb}
\usepackage{siunitx}

\usepackage{color}

\draft % marks overfull lines with a black rule on the right

\begin{document}

%\preprint{}

\title{Power stabilization of a diode laser with an acousto-optic modulator  } 



\author{F. Tricot}
\email[]{francois.tricot@obspm.fr}
\affiliation{LNE-SYRTE, Observatoire de Paris, Universit\'{e} PSL, CNRS, Sorbonne Universit\'{e}, 61 avenue de l'Observatoire, 75014 Paris, France.}
\affiliation{ Thales AVS, 2 rue Marcel Dassault, 78140 Velizy-Villacoublay, France.}
\author{D. H. Phung}
\affiliation{LNE-SYRTE, Observatoire de Paris, Universit\'{e} PSL, CNRS, Sorbonne Universit\'{e}, 61 avenue de l'Observatoire, 75014 Paris, France.}
\author{M. Lours}
\affiliation{LNE-SYRTE, Observatoire de Paris, Universit\'{e} PSL, CNRS, Sorbonne Universit\'{e}, 61 avenue de l'Observatoire, 75014 Paris, France.}
\author{S. Gu\'erandel}
\affiliation{LNE-SYRTE, Observatoire de Paris, Universit\'{e} PSL, CNRS, Sorbonne Universit\'{e}, 61 avenue de l'Observatoire, 75014 Paris, France.}
\author{E. de Clercq}
\affiliation{LNE-SYRTE, Observatoire de Paris, Universit\'{e} PSL, CNRS, Sorbonne Universit\'{e}, 61 avenue de l'Observatoire, 75014 Paris, France.}

%\affiliation{LNE-SYRTE, Observatoire de Paris, Universit\'{e} PSL, CNRS, Sorbonne Universit\'{e}, 61 avenue de l'Observatoire, 75014 Paris, France}

%\email{neumeier@physics.montana.edu}
%\thanks{}

%\noaffiliation is required (may also be used with the \author command).

%\date{\today}

\begin{abstract}   % Abstract goes here.

Laser power fluctuations can significantly reduce the device performances in various applications. High frequency fluctuations impact the signal-to-noise ratio, while slow variations can reduce the device repeatability or accuracy. Here we report experimental investigations on the power stabilization of a diode laser with an acousto-optic modulator. In the frequency domain 
the relative power noise is reduced at the level of $2.2\times 10^{-8}$ Hz$^{-1/2}$ in the range 1-100 kHz. The slow variations are studied in the time domain. The relative Allan standard deviation is measured at the level of $6\times 10^{-7}$ at 100 s averaging time. Above 100 s the instability increases and reaches $2 \times 10^{-6}$ at 10 000 s.

\end{abstract}

% \pacs{07.07.Tw, 42.55.Px, 42.60.Mi, 42.79.Jq}% insert suggested PACS numbers in braces on next line
%{07.07.Tw, 42.55.Px, 42.60.Mi,42.79.Jq} 
% 07.07.Tw Servo and control equipment; robots
% 42.55.Px 	Semiconductor lasers; laser diodes / 42.60.Mi 42.60.Mi 	Dynamical laser instabilities; noisy laser behavior / 42.79.Jq 	Acousto-optical devices
\maketitle %\maketitle must follow title, authors, abstract and \pacs

% References should be done using the \cite and \label commands.
 
\section{Introduction}

Stable laser powers are needed in a wide range of  applications such as laser writing systems \cite{Kim:RSI:2007}, magnetometers \citep{Duan:OE:2015}, atomic clocks \cite{Lin:FCS:2014,Kozlova:TIM:2014,Abdel:JAP:2017}, spectroscopy \cite{Gehrtz:JOSAB:1985,DuBurk:MST:2004, Casa:JCP:2007,Moretti:PRL:2013}, laser frequency standards, interferometry and gravitational wave detection \citep{Kwee:OE:2012,Junker:OL:2017}, etc. High frequency fluctuations of laser power decreases the signal-to-noise ratio, degrading the short-term frequency stability of frequency standards \cite{Abdel:JAP:2017}, or can distort line-shapes in frequency modulation spectroscopy \citep{DuBurk:MST:2004}. Low frequency variations can degrade long-term stability of atomic clocks \cite{Kozlova:TIM:2014,Abdel:JAP:2017}. 
Various approaches can be employed to tackle this issue. Aside from passive isolation \citep{Talvitie:RSI:1997}, the optical power can be stabilized by a feedback on the diode current (at the expense of the frequency stability) or temperature \cite{Lee:RSI:1996}.  Most often  an external actuator is used for this purpose, like an electro-optic modulator \citep{ Ivanov:UFFC:2009, Liu:OLT:2013}, a photo-elastic modulator \citep{Duan:OE:2015}, or an acousto-optic modulator (AOM) \citep{Kim:RSI:2007, Casa:JCP:2007,Moretti:PRL:2013,Kwee:OE:2012, Junker:OL:2017,Tetchewo:FNL:2007,Lin:FCS:2014,Balakshy:OLT:2014}. A great deal of work has been devoted for decades to the AC power stabilization of Nd:YAG lasers for gravitational wave detectors. A relative power noise (RPN) at the outstanding level of $1.8\times 10^{-9}$ Hz$^{-1/2}$ in the 100~Hz-1~kHz band has recently been demonstrated \citep{Junker:OL:2017} with AOM.
 However these highly sophisticated devices are too complex for industrial applications or ordinary laboratories, and cannot be implemented in on-board devices such as compact atomic clocks which must be small and low-power. Although  the laser-power stabilization by means of an AOM is well known and very common in many laboratories, surprisingly, it is very poorly documented in the literature. Particularly the slow power fluctuations and drifts are not addressed. 
 
 The aim of this paper is to fully describe and characterize a simple and efficient setup for on-board applications or ordinary laboratories, so that it could be used as a guideline. The limiting noise sources of the device and of the measurement are carefully addressed. The device performances are characterized in the frequency domain (power spectral density (PSD) in 1 Hz-10 MHz band) and in time domain (very slow frequencies, averaging times from 1 s to 10 000 s). 
  They are investigated by means of the Allan standard deviation, which characterizes the power instability as a function of the averaging time. RPN and Allan deviation both allow an analysis of the limiting noise types, but in frequency domain and time domain, respectively. The link between noise power spectral density and Allan deviation, well known in time and frequency metrology \citep{Rubiola:PN:2008}, will be used to confirm the analysis.

In this paper we characterize the power stabilization of a diode laser based on a simple scheme using a single acousto-optic modulator (AOM). This is needed for a project of compact vapor-cell atomic clock based on coherent population trapping. Vapor cell atomic clocks based on coherent population trapping (CPT) \cite{Arimondo:1996, Kitching:TIM:2000, Merimaa:JOSAB:2003, Vanier:APB:2005} are promising devices for their potential compact size \cite{ShahKitching:2010} or their  high performances \cite{Danet:UFFC:2014, Abdel:JAP:2015, Yun:PRAp:2017, Abdel:JAP:2017}. In such clocks  there is no microwave cavity unlike well-known Rb vapor cell clocks \cite{Vanier:TIM:2003}, the microwave signal is optically carried by a bi-frequency laser beam. The CPT clock resonance occurs when the frequency difference between both optical frequencies is equal to the microwave clock frequency. The resonance is detected by recording the optical power of the light transmitted through the vapor cell. The counterpart of this all-optical interrogation and detection is the high sensitivity to the laser power. The signal amplitude and the clock frequency are both power sensitive. 

\begin{figure}[htbp]
\centering
\includegraphics[width=\linewidth]{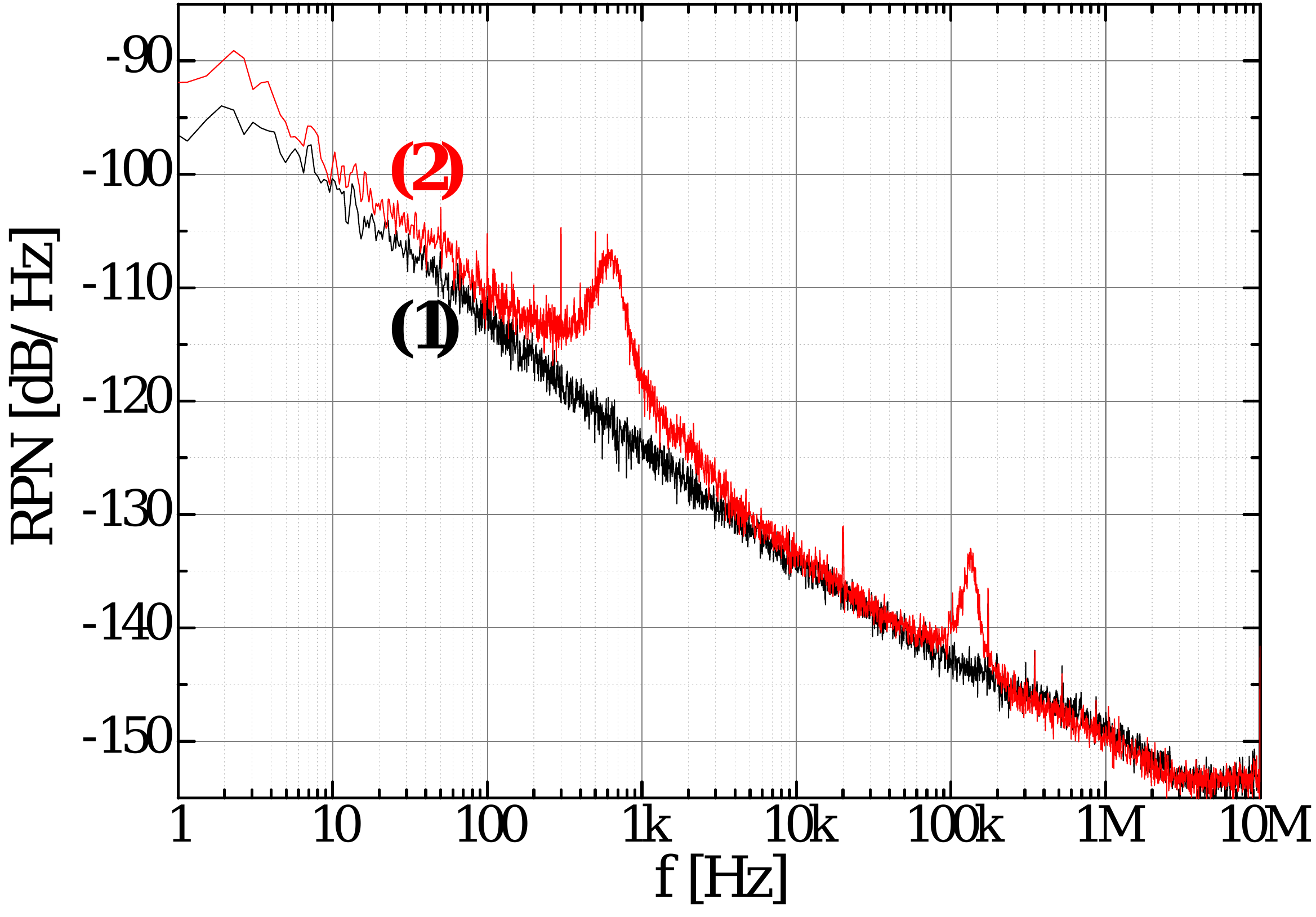}
\caption{Relative power noise of the ECDL laser. (1) Free running; (2) the laser frequency is locked on the Cs D1 transition.}
\label{fig:RINfreq}
\end{figure}
In our setup the two frequencies are produced by two home-made extended-cavity diode lasers (ECDL) \cite{Baillard:2006}. One laser is frequency stabilized by saturated absorption spectroscopy on the Cs D1 line at 895 nm.
% The second laser is locked on the first one with a 9 GHz offset by an optical phase-lock loop. In both lasers 
The slow frequency corrections are applied on a piezoelectric transducer (PZT) controlling the cavity length, while fast corrections are applied on the diode driving current. Such actions, cavity-length and current variations, in turn modify the output power and alter the power stability. 
The PSD of the relative power noise (RPN, also called RIN in the literature) of a diode laser is shown in Fig. \ref{fig:RINfreq}. As expected the noise spectrum of the free-running laser is dominated by flicker noise. The laser frequency locking adds two supplementary noise peaks on the spectrum, related to the PZT and diode current at low and high frequency, respectively.

Since the  laser power fluctuations are one of the main sources of frequency instability in our clock, we implemented a simple low noise power lock for each laser based on AOMs. Fast fluctuations affect the  short-term frequency instability, and are characterized by the RPN. Slow power variations are one of the major  source of instability at midterm averaging times. They are investigated by means of the Allan standard deviation.

The paper is organized as follows. We first describe the experimental set-up  controlling the laser power and the servo electronic circuit. The results on the RPN reduction, in loop and out-of-loop, are then presented. The last part is devoted to the analysis and reduction of the slow power fluctuations. 

\section{Power servo}

\begin{figure}[htbp]
\centering
\includegraphics[width=\linewidth]{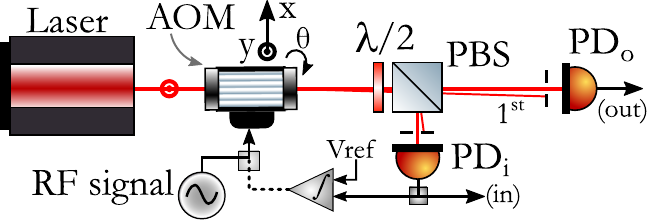}
\caption{Experimental set-up stabilizing the laser power. The power lock controls the level of the RF signal. AOM acousto-optic modulator, $\lambda/2$ half-wave plate, PBS polarizing beam splitter, V$_{\mathrm{ ref}}$ voltage reference,  PD$_{i}$ (PD$_{o}$) in-loop (out-of-loop) photodiode, respectively .}
\label{fig:manip}
\end{figure}

A schematic of the setup used for power locking is shown in Fig. \ref{fig:manip}. The laser beam passes through an AOM (Crystal Technology 3080) in the Bragg configuration. The AOM position is set to optimize the first-order diffraction beam thanks to a xy dual-axis translation stage and a  $\theta$ rotation mount (Fig. \ref{fig:manip}). The first-order beam is blocked further away. Usually, a part of the zero-order beam is picked up by using a beamsplitter of split ratio  10/90 or 30/70.  For this experiment, the beamsplitter is replaced by an half-wave plate and a polarizing cube (PBS) to get a variable split ratio. The transmitted beam is the useful beam, its power $P_{out}$ is monitored by a photodiode PD$_o$ (out-of-loop signal), the reflected beam power ($P_{in}$) is detected by an identical photodiode PD$_i$ (Thorlabs-PDA36A), yielding the in-loop signal. The optical bench is enclosed in a box whose temperature stability is of several tens of mK.

The in-loop signal is compared to a reference voltage issued from a low noise and low drift precision voltage reference (Linear Technology, LT1021-10V). The error signal, difference between the photodiode signal and the reference, is  integrated through a proportional-integrator controller (PI). The correction is applied on an attenuator (Mini-Circuit, TFAS-2+) controlling the power level of the 80 MHz radio-frequency (RF) signal driving the AOM, i.e. the servo controls the power balance between AOM diffraction orders. 
The control efficiency is characterized in the frequency domain by the noise PSD. For Fourier frequency above 1 Hz  the noise spectra of the various signals are recorded by a fast-Fourier-transform analyzer (FFT, Agilent-89410A). In the time domain, for averaging time above 1~s, the slow variations are characterized by the Allan standard deviation. A data acquisition unit (Agilent 34970) is used for time measurements of the laser power.

\begin{figure}[htbp]
\centering
\includegraphics[width=\linewidth]{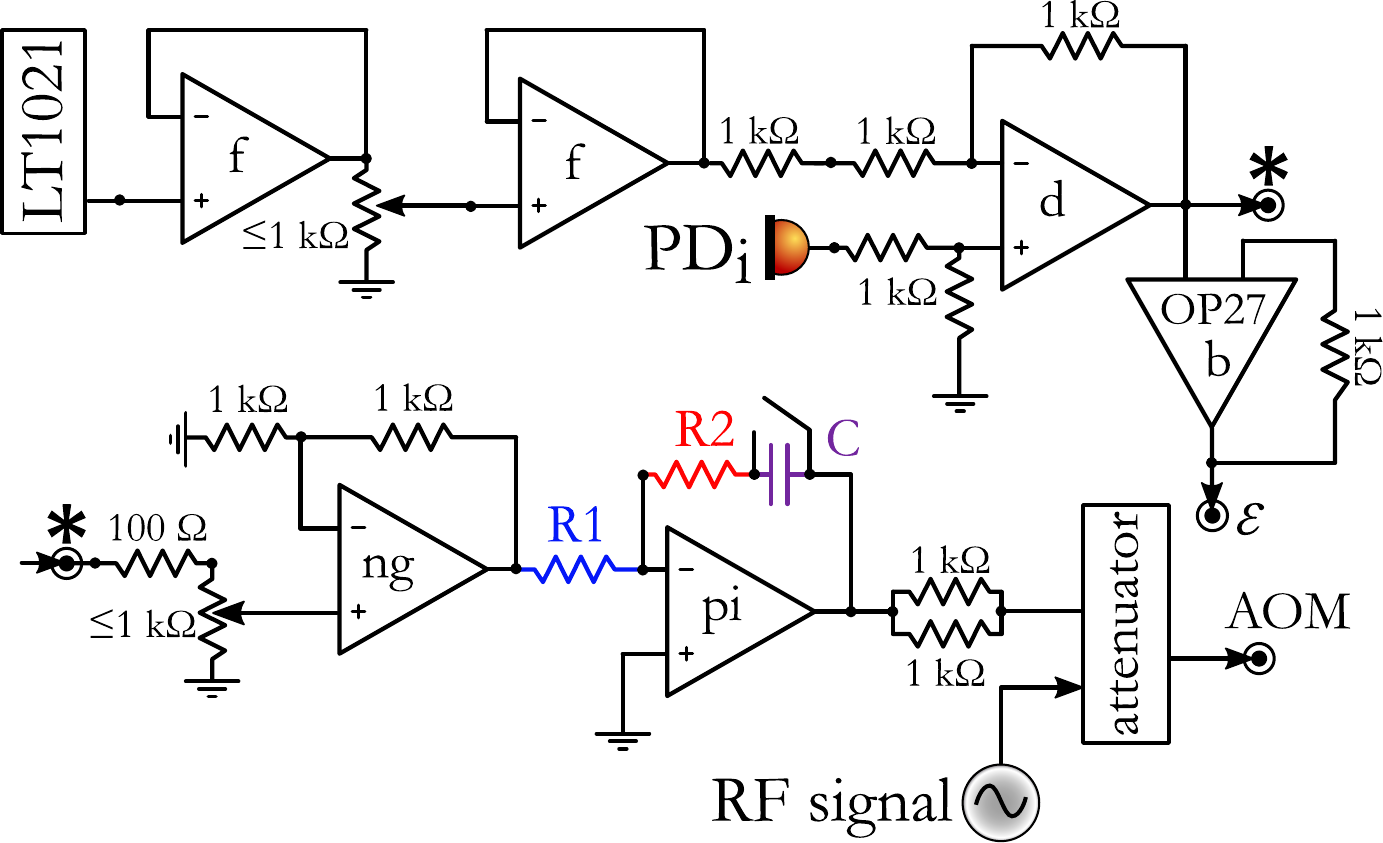}
\caption{Power-lock circuit. LT1021,  high-stability voltage reference. The OP amplifiers are MAX9632 except for the OP27 buffer b, f unity gain buffer amplifier, d differential amplifier, ng non-inverting gain amplifier, pi proportional-integrator controller, PD$_{i}$ in-loop photodiode.}
\label{fig:circuit}
\end{figure}

The electronic diagram is shown in Fig. \ref{fig:circuit}. 
%The voltage reference \textbf{ amplification} is optimized by using 
Low-noise and wide-band operational amplifiers (OP) MAX9632 are used, 
except for the buffer amplifier (b). Two unity gain buffer amplifiers (f) isolate possible noise returns which would degrade the characteristics of the voltage reference. In addition, the second OP (f) is used to adjust the voltage reference at the desired level. 
This adjustment is made by monitoring the error signal $\epsilon$ available with the OP27 buffer (b). After a non-inverting gain amplifier (ng), the error signal is processed  by a PI controller with a bandwidth of about 700~kHz. The optimal parameters to reduce the RPN laser close to the relative voltage reference are founded to be R1~=~5~k\ohm, R2~=~1~k\ohm~and C~=~220~pF. 

\section{Relative power noise}
\subsection{In-loop power value }
\label{subsec:inloop}

\begin{figure}[t]
\centering
\includegraphics[width=\linewidth]{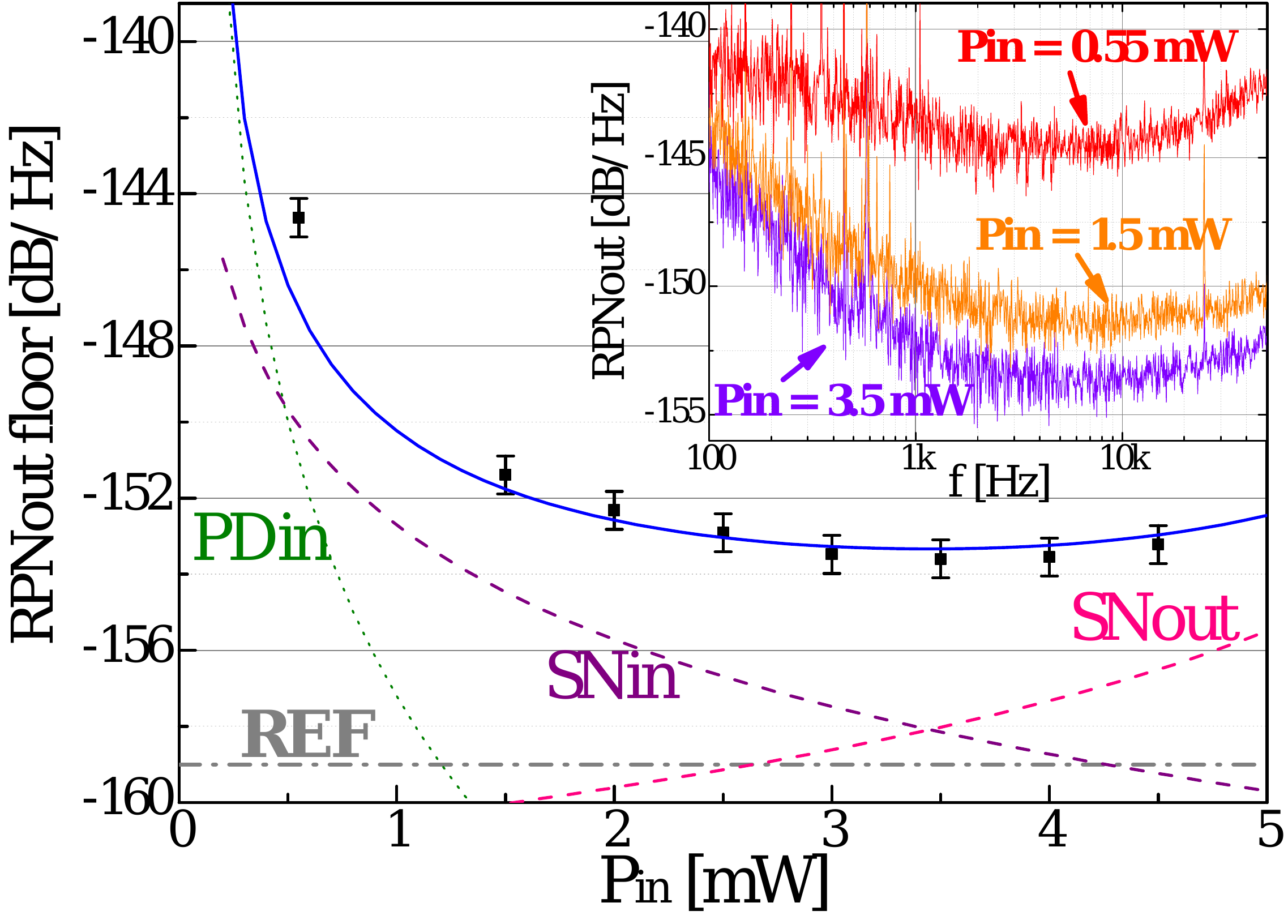}
\caption{Noise floor value of the out-of-loop RPN as a function of the in-loop power laser $P_{in}$. Black squares experimental data, blue solid line computed value. The floor value is the RPN level at 4 kHz Fourier frequency. REF, PDin, SNin, and SNout contributions of the voltage reference, in-loop photodiode, in-loop and out-of-loop shot noises, respectively.  The inset shows the out-of-loop RPN measured from 100~Hz to 50~kHz for three values of $P_{in}$.}
\label{fig:RINvsPin} 
\end{figure}

The laser power $P_{in}$ used in the servo loop must be high enough so that photodiode noise and shot noise are negligible. However, in our set-up the total power available at the output of the AOM is low, only about 7~mW. As the $P_{in}$ value increases, the out-of-loop power $P_{out}$ decreases, so that its shot-noise increases in relative value.In this way the value of $P_{in}$ minimizing the in-loop noise, is not the value minimizing the out-of-loop noise floor. With the half-wave plate and the PBS we experimented different values of $P_{in}$ to reach the lowest out-of-loop RPN floor. The experimental results are shown in Fig. \ref{fig:RINvsPin}. The out-of-loop RPN noise floor value is taken at 4~kHz Fourier frequency from the RPN measurements presented in the inset. The experimental data (squares) are in fairly good agreement with a simple estimation (solid line) given by 
%\begin{equation}
$RPN_{out} \sim REF+ PD_{in}+ PD_{out}+SN_{in}+SN_{out} $,
%\label{eqn:RINout}
%\end{equation}
with $REF$ the voltage-reference noise PSD in relative unit,  $SN_{in}$ and $SN_{out}$ the relative shot-noise PSDs of the in-loop and out-of-loop signals, respectively. $PD_{in}$ and $PD_{out}$ are the noise PSDs of the photodiodes PD$_{i}$ and PD$_{o}$ measured in the dark and divided by the squared mean signal value.
 
At low $P_{in}$ values, the power lock cannot efficiently filter the laser noise because the in-loop noise is dominated by the photodiode noise. When  $P_{in}$ is increased above 4 mW, in order to reduce the in-loop photodiode noise contribution, the out-of-loop power is decreased. Therefore the out-of-loop RPN floor value is dominated by the shot-noise of the out-of-loop signal. In the following, $P_{in}=3$~mW, this value is a good trade-off to avoid the contribution of the in-loop detection noise and to keep enough power laser for the useful beam.

\subsection{In-and out-of-loop results}

\begin{figure}[t]
\centering
\includegraphics[width=\linewidth]{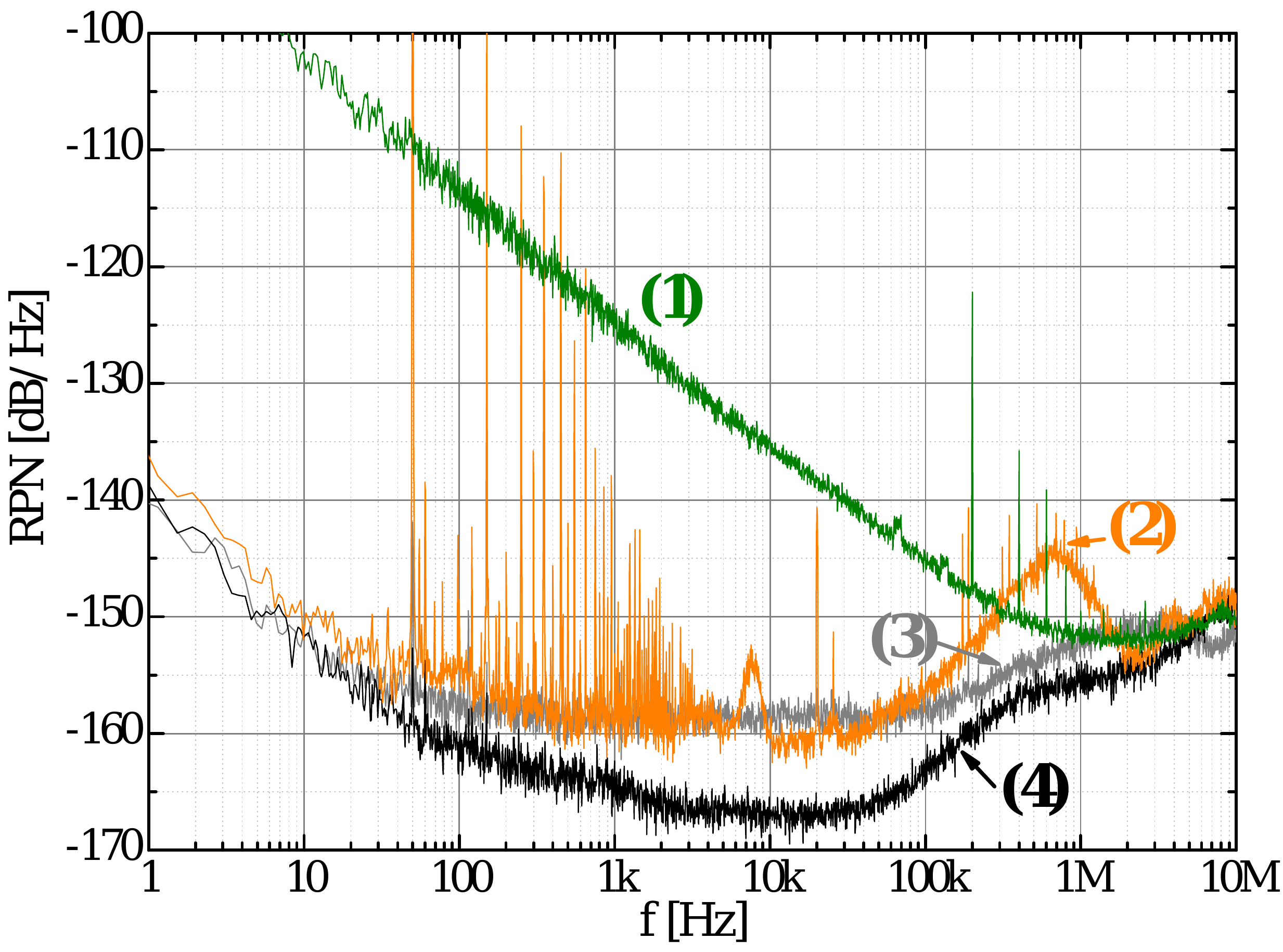}
\caption{ In-loop RPN in free running (1) and power-locked (2) regimes. (3) Normalized voltage-reference noise, (4) photodiode noise. }
\label{fig:RINinloop}
\end{figure}

The in-loop measurement of the RPN PSD is shown in Fig. \ref{fig:RINinloop}, together with the noise PSDs of the voltage reference and of the photodiode. The voltage reference noise is measured after the differential amplifier (d) (Fig. \ref{fig:circuit}). For purposes of comparison all PSD are normalized to the photodiode-signal mean value.
When the servo-loop is closed, the RPN is reduced by 40 dB  at 100~Hz Fourier frequency. From 200~Hz to 60~kHz the noise floor is close to the floor of the voltage reference at the level of $-158.5$~dB/Hz. The bump visible on the RPN curve around 600~kHz Fourier frequency does not exceed $-144$~dB/Hz. Above 2~MHz, the measurement is limited by the photodiode noise. This result shows that the servo loop can filter the laser-power noise in the loop at the level of the reference noise up to 100 kHz. The same result is obtained on both laser systems.

\begin{figure}[htbp]
\centering
\includegraphics[width=\linewidth]{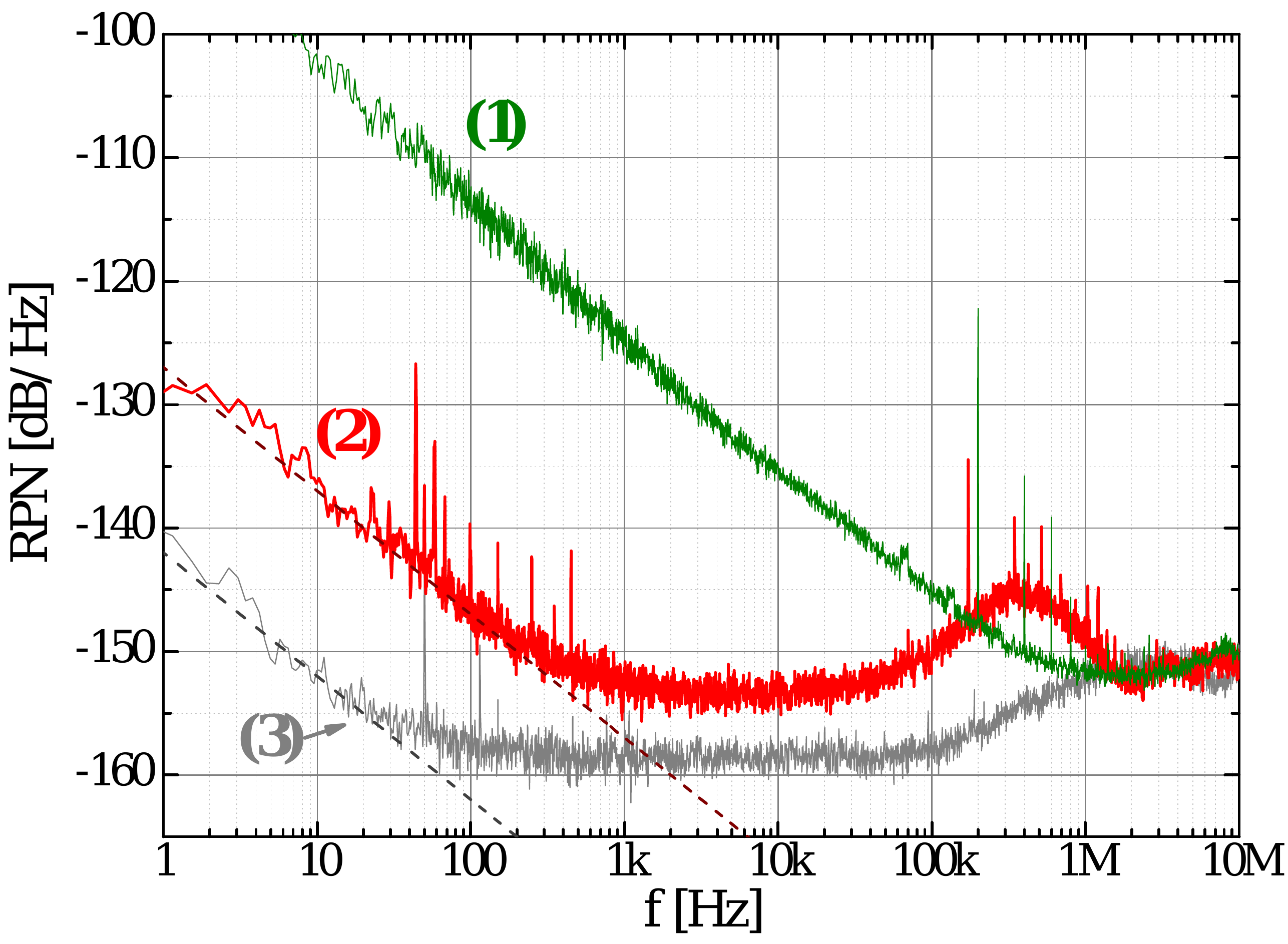}
\caption{Out-of-loop RPN in free running (1) and power-locked (2) regimes. (3) Normalized voltage-reference noise. The dashed lines represent the asymptotic flicker noise at low frequency.}
\label{fig:RINoutloop}
\end{figure}

The out-of-loop  RPN is shown in Fig. \ref{fig:RINoutloop}.
When the power servo-loop is locked the RPN level is about 
$-147$~dB/Hz at 100~Hz Fourier frequency, a reduction of about 32 dB compared to the unlocked power case. The noise floor level from 1~kHz to 100~kHz is limited by the shot noise at the level of 
$2.2\times 10^{-8}$ Hz$^{-1/2}$. Then the bump servo loop is visible around 500~kHz Fourier frequency.
 The asymptotic behavior in $h_{-1}f^{-1}$ of the flicker noise at low Fourier frequencies $f$ will help to understand the Allan deviation measurements in the next section. We recall \citep{Rubiola:PN:2008} that a PSD of function $h_{-1}f^{-1}$ ($h_{0}$) leads to an  Allan deviation equals to $\sqrt{2 \ln(2)h_{-1}}$  ($\sqrt{h_0/2}\tau^{-1/2}$), respectively, where $\tau$ is the averaging time.
 For the  voltage reference and for the out-of-loop RPN  we get: $h_{-1}=6.3\times10^{-15}$ ($-142$~dB) and 
$h_{-1}=2\times10^{-13}$ ($-127$~dB), respectively.

This noise level leads to a reduction of the RPN contribution to the frequency instability of our CPT clock at the level of 2.5$\times10^{-14}$ at 1~s averaging time instead of 5$\times10^{-13}$ when laser powers are not stabilized.

\section{Slow Power fluctuations}
\subsection{Data acquisition characterization.}

\begin{figure}[htbp]
\centering
\includegraphics[width=\linewidth]{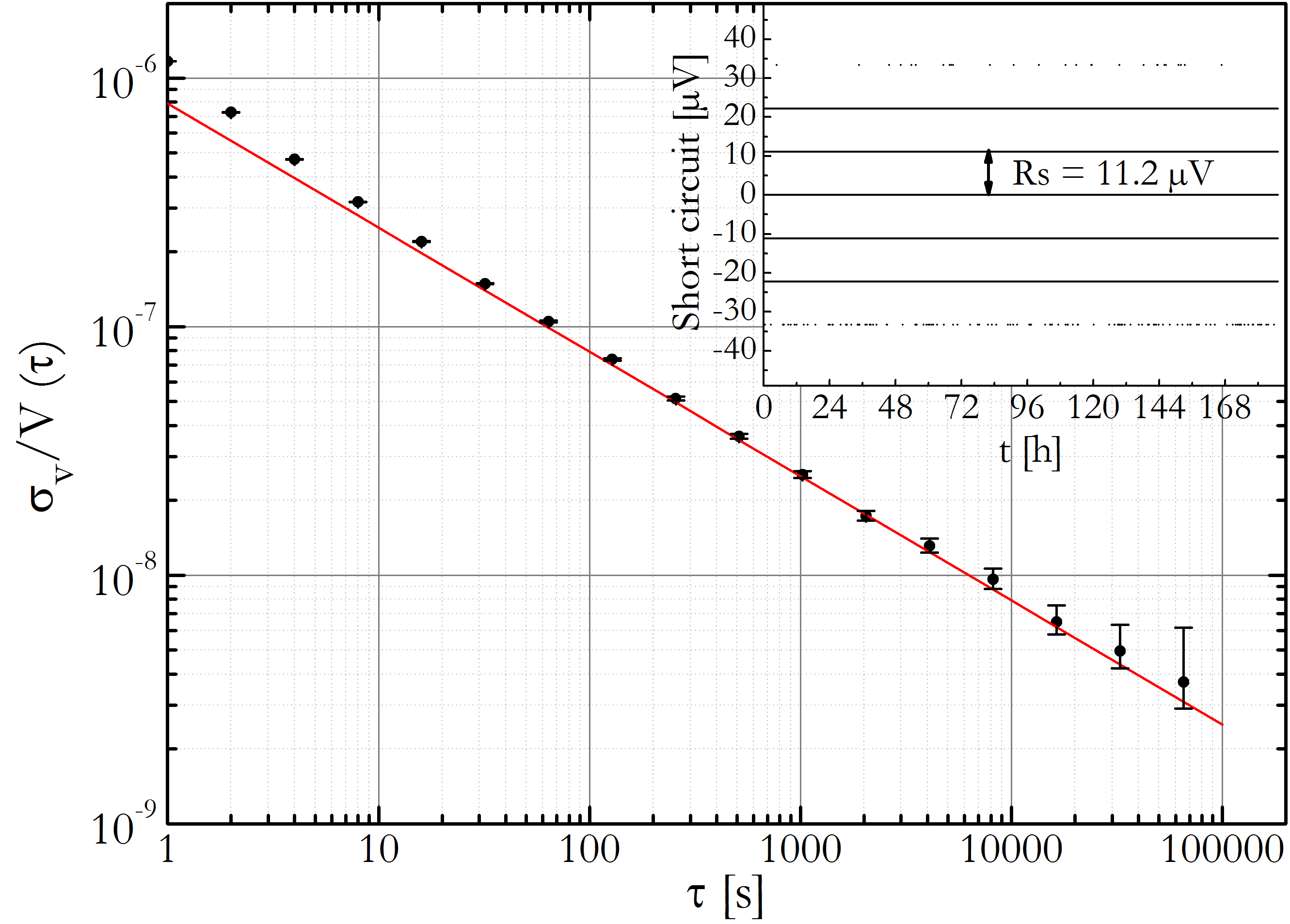}
\caption{Allan deviation of the voltage noise measured on a short-circuit. The deviation is normalized to  10~V. Dots data, red line computed line of slope $\sqrt{r^2/2}\tau^{-1/2}$. The inset shows the measurement during 180~h.}
\label{fig:Agilent}
\end{figure}

Before investigating the slow power fluctuations we checked the possible contributions of the voltage measuring instrument and of the voltage reference. We measured the low-frequency fluctuations of the voltage reference and of the laser power by using the data logger. As we needed to measure fractional laser-power fluctuations  below $\sigma_P/P\sim 1\times10^{-6}$ at $10\,000$~s integration time, we first characterized the data logger stability. This instrument has a typical resolution of about \SI{10}{\micro\volt} for a 10~V range. The result of a measurement of the null voltage of a short-circuit during more than 180~hours , is shown in inset of Fig. \ref{fig:Agilent}. The measured resolution is $R_S=11.2$ \SI{}{\micro\volt}. The measurement stability is characterized by the Allan standard deviation (Fig. \ref{fig:Agilent}) which is here divided by 10~V for purposes of comparison with further power measurements in fractional units. The slope in $\tau^{-1/2}$, with $\tau$ the averaging time, is the feature of a white noise. The slope here is equal to  $(r/\sqrt{2}) \tau^{-1/2}$, with $r=R_S/10$.
This result shows that the data logger can measure relative fluctuations below 1$\times10^{-6}$ at $10\,000$~s averaging time.

\subsection{Voltage reference instability}
\begin{figure}[htbp]
\centering
\includegraphics[width=\linewidth]{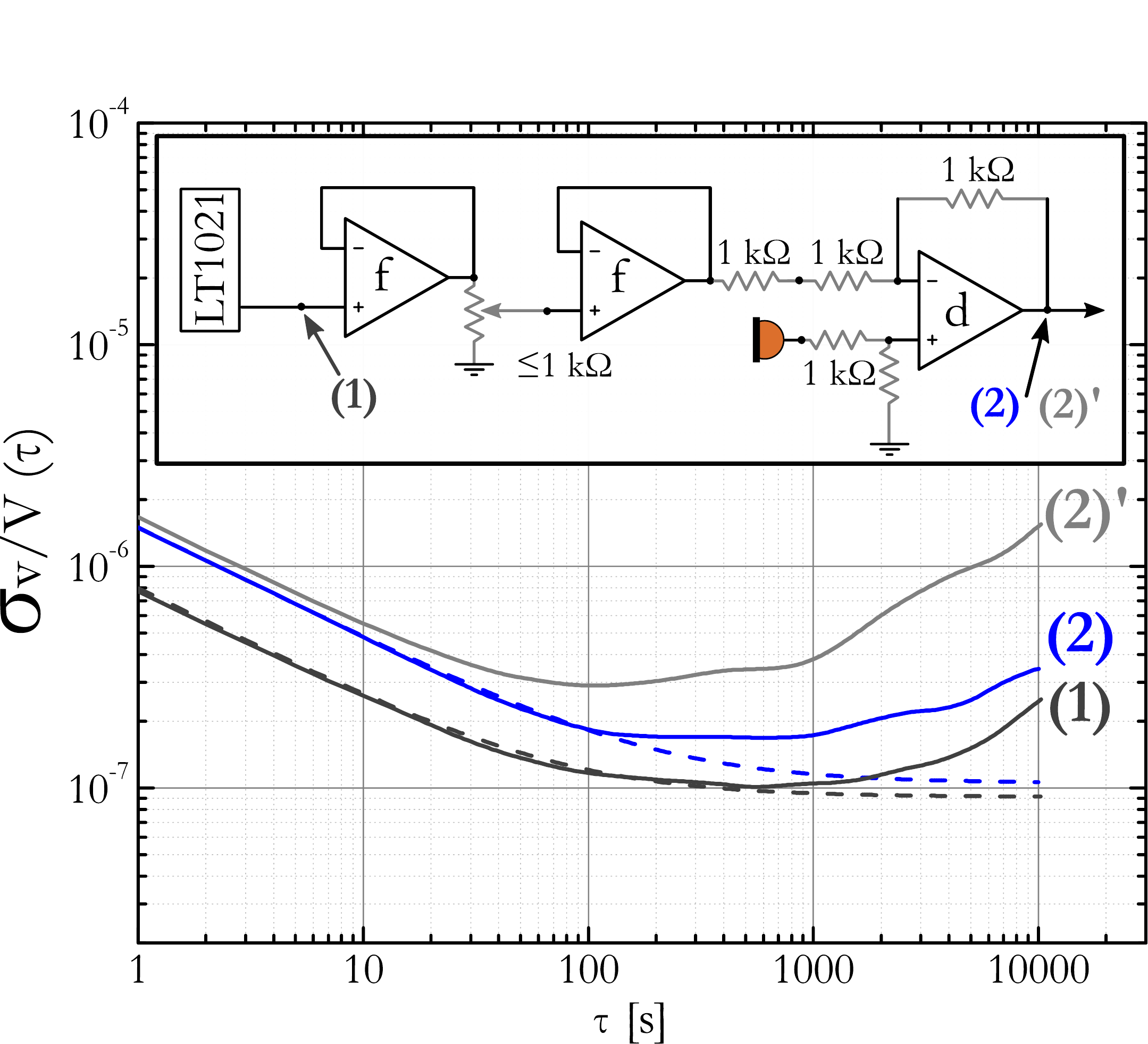}
\caption{Allan deviation of the relative voltage reference measured at (1) LT1021 output (10~V), and (2) after the differential amplifier (5~V) with electronic board using 
low temperature coefficient resistors or common resistors (2'). The dashed line represent the expected results by summing only contributions of data-logger white noise and LT1021 flicker noise. The inset scheme shows where the measurements are performed.}
\label{fig:vishay}
\end{figure}

We measured with the data logger the reference voltage at various points of the power lock circuit (see inset of Fig. \ref{fig:vishay}), (1) at the output of the LT1021 reference, (2) at the output of the differential amplifier with the in-loop photodiode PD$_i$ in the dark.  
The measured relative Allan deviations are shown in Fig. \ref{fig:vishay} up to $10\,000$~s averaging time. Note that the values are normalized to 10~V for curve (1), and 5~V for curve (2), which is the mean value in working conditions.  The curve (2') shows the measured instability with a previous version of the electronic board using common resistors (temperature sensitivity 5~ppm/K) instead of low temperature coefficient (Vishay) resistors (0.05~ppm/K).

At short term ($\tau< 100$ s) the measurement of the LT1021 instability (curve (1)) is limited by the data logger noise. 
The LT1021 flicker noise (see Fig. \ref{fig:RINoutloop}) yields indeed a relative Allan deviation floor of about $\sqrt{2 \ln(2)h_{-1}}=1\times 10^{-7}$, well below the data logger deviation. The voltage reference floor is reached between 100 and $1\,000$~s averaging times.

 The instability begins to increase above $1\,000$~s. The instability of the reference voltage measured after the differential amplifier (curve (2)) has the same behavior. The factor of 2 compared to curve (1) is due to the different dividing values (5~V and 10~V) for the same noise value.
 In the following the power instability will be compared to the minimum measurable instability  given by  $\sigma^2(\tau)= (r^2/2) \tau^{-1}+ 2\ln(2)h_{-1}$.

\subsection{Laser Power instability}

\begin{figure}[htbp]
\centering
\includegraphics[width=\linewidth]{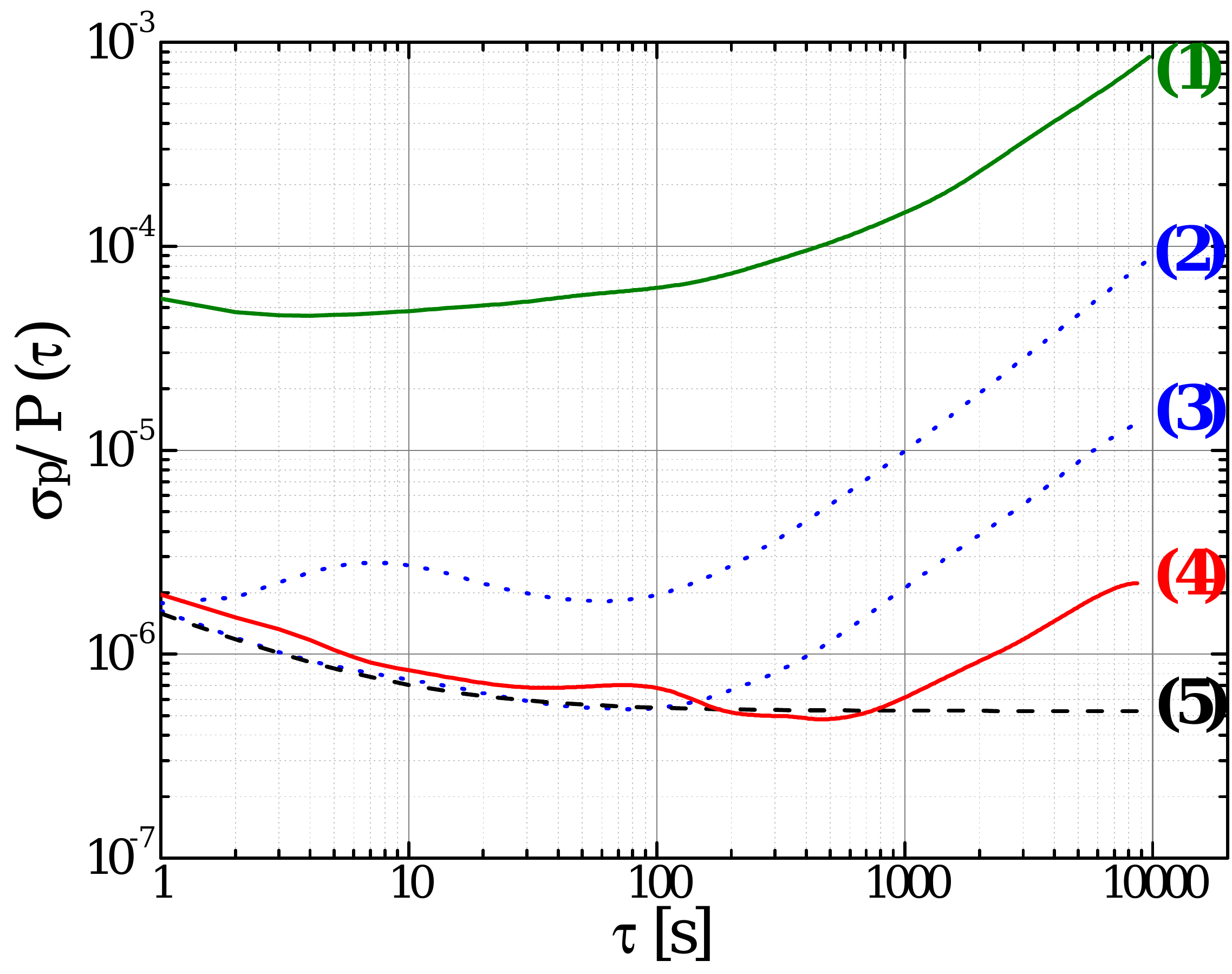}
\caption{Relative laser-power instability. (1) Free-running power, (2) power-locked regime without AOM temperature regulation, (3)  power-locked regime with  AOM temperature regulated at 27.8~\celsius, (4) power-locked regime with  AOM temperature regulated at 27.8~\celsius, and with the use of an uncoated BS to pick up the laser beam, (5) expected instability.}
\label{fig:StabP}
\end{figure}

The measured laser-power instability of a single laser is shown in Fig. \ref{fig:StabP}. When the power loop is closed, the instability (2) is one order of magnitude lower than for free-running regime (1). 
According to the flicker noise level in Fig. \ref{fig:RINoutloop}, at short averaging times the closed-loop instability should be $5.3 \times 10^{-7}$. We cannot measure this level masked by the data logger instability. 
At 1~s averaging time the instability is closed to the expected value (5) computed by summing theoretical contributions of data logger and RPN. However, after 1~s, the instability increases  and reaches 9$\times10^{-5}$ at $10\,000$~s. The Allan deviation scales as $\tau$, which is consistent with a linear drift. 
This drift cannot be explained by the voltage reference whose contribution is here negligible (see Fig. \ref{fig:vishay}).

\begin{figure}[htbp]
\centering
\includegraphics[width=\linewidth]{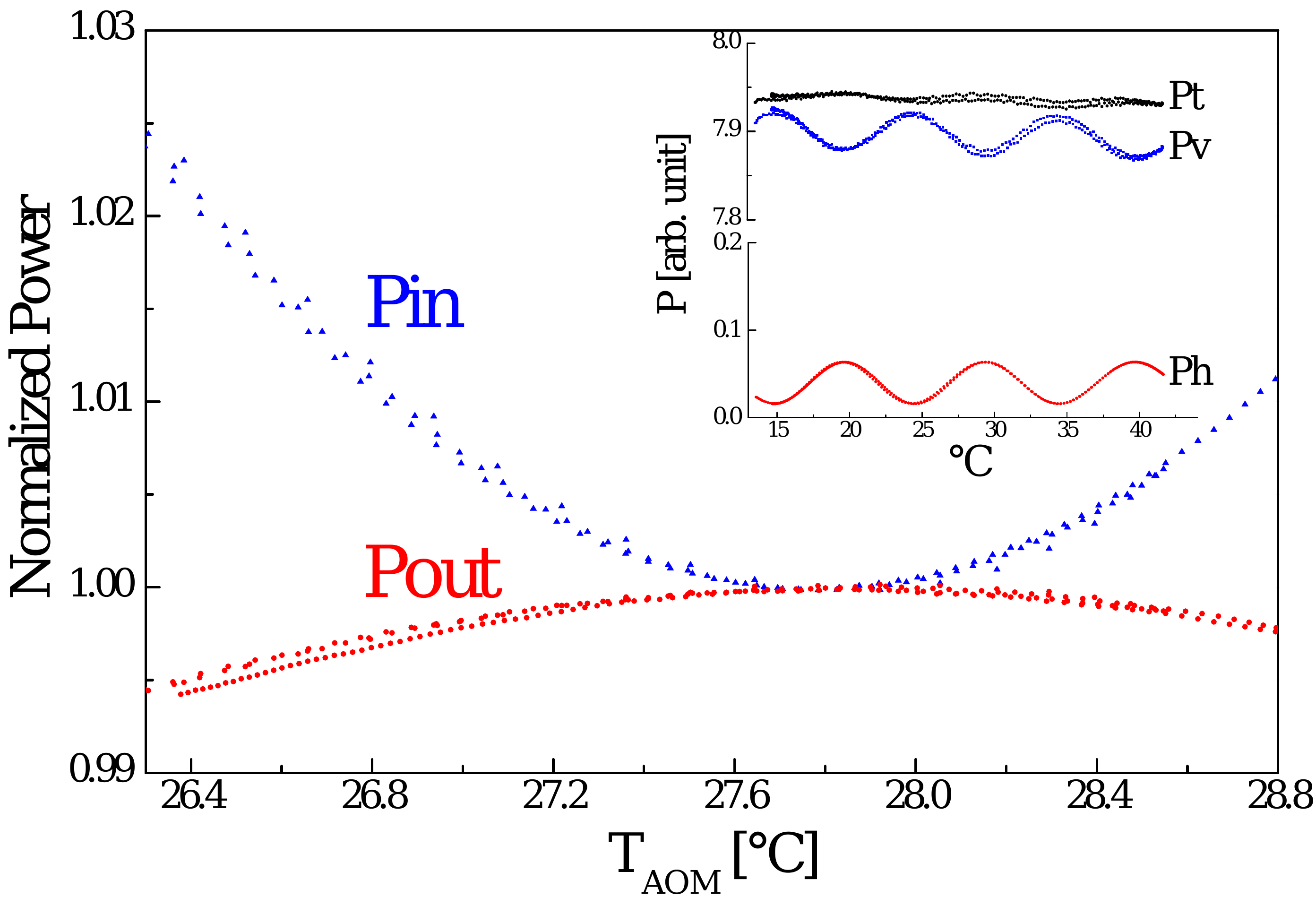}
\caption{Normalised optical powers measured in and out of the loop as a function of the AOM temperature. Each power is normalized to its mean value at 27.8~$\celsius$. The data scattering is due to back and forth scans. Inset: Powers measured at the PBS output for a linearly or vertically polarized beam at the AOM input. Ph: horizontal polarization, Pv: vertical polarization, Pt: sum of Ph and Pv powers. }
\label{fig:PvsTaom}
\end{figure}

Actually, the power variations of $P_{out}$ are correlated to the temperature variations in the box enclosing the experimental set-up. We identified the AOM as one of the main temperature sensitive component ($\sim 10^{-2}$/K). The AOM crystal is a tellurium dioxide (TeO$_2$ or paratellurite) crystal, known to be highly birefringent \citep{Ohmachi:1970,Uchida:1976, Balakshy:1996}. Then if the incident polarization is not perfectly linear and aligned on one crystal axis the output polarization will be rotated or its ellipticity will change, and this effect is temperature sensitive. 
 To characterise it, we have recorded the powers $P_{in}$ and $P_{out}$ as a function of the AOM temperature, regulated by a thermo-electric cooler (TEC) placed under it.
 The inset in figure \ref{fig:PvsTaom} shows the powers measured at the output of the PBS for the horizontal and the vertical  polarizations. The beam at the input of the AOM is linearly and vertically polarized, and the half-wave plate in-front of the PBS is removed. The vertical polarization is the polarization recommended by the manufacturer for the best diffraction efficiency. A polarization modulation is clearly observed. Note that the same modulation is observed with the horizontal polarization with a smaller amplitude.       
 Figure \ref{fig:PvsTaom} shows the normalized in-and-out-loop powers around 28~$\celsius$ in working conditions (the plate is in front of the PBS, and tilted for tuning the split ratio). $P_{in}$ ($P_{out}$) is the power of the vertically (horizontally) polarized output beam, respectively. 
        It is clear that the AOM is sensitive to temperature variation, in such a way that the AOM crystal changes the polarization of the output laser beam before the PBS. Consequently, when the power loop is closed, a temperature variation leads to opposite variation of $P_{in}$ and $P_{out}$. The servo regulates $P_{in}$ to a constant value by correcting the total power by means of the RF power driving the AOM, and so the $P_{in}$ lock increases the variation of $P_{out}$. 
%This explains the difference between measured in-loop and out-of-loop instabilities. 

 An interesting point to be noticed is the AOM temperature sensitivity around 27.8~$\celsius$, which shows an inversion temperature (Fig. \ref{fig:PvsTaom}). To confirm this effect, we measured the power instability out of the loop while the AOM temperature is regulated at 27.8~$\celsius$ by a proportional-integral-derivative (PID) controller. The result is shown in Fig. \ref{fig:StabP}(3). We clearly see an improvement of the stability, now equal to the expected value (5) until 100~s. Between 100 and 10 000 s an improvement of a factor close to five  is obtained.
This result confirms the AOM temperature influence on the power instability. 
 Nevertheless, a power drift is still visible above 500~s. The Allan deviation of the regulated AOM temperature is $6\times 10^{-5}$ $\celsius$ and $1\times 10^{-3}$ $\celsius$  at averaging times of 100 and $10\,000$~s, respectively. These values lead to fractional out-of-loop power deviations of $4\times 10^{-8}$ and $6\times 10^{-7}$. The remaining power drift is then not due to the AOM sensitivity. 
In order to overcome the birefringent effect in the AOM, we inserted a polarizer between the AOM and the half-wave plate, so that the polarization does not change at the input of the plate. This did not reduce the observed power drift. As a matter of fact, the residual thermal contribution is due to the use of the half-wave plate and the PBS, which was used to tune the beam part extracted for the servo-loop. Its sensitivity has been measured to be about $\leq8\times10^{-3}$/K. We have then replaced the plate-PBS set by 
an uncoated glass plate. As seen in section \ref{subsec:inloop} the in-loop power is no more optimized for PSD floor, but here the purpose is to check the impact on long term variations. The resulting relative power instability is shown in Fig. \ref{fig:StabP}. An improvement of a factor six is obtained at $10\,000$~s, at the level of  2.5$\times 10^{-6}$ (Fig. \ref{fig:StabP}(4)). The residual power drift measured from $1\,000$~s to $10\,000$~s is attributed to  temperature sensitivity of both photodiodes ($\sim$~2$\times 10^{-4}$/K).
Note that the beam pointing stability after an AOM is also sensitive to the RF power \citep{Frohlich:RSI:2007}. Here the fractional fluctuations of the RF power is less than $10^{-3}$ leading to an angular stability below \SI{1}{\micro rad} with data of Ref.\citep{Frohlich:RSI:2007}. As the active area of the photodiodes is 13 mm$^2$ we can neglect such an angular variation.

\section{Conclusion}

We have reported the characterization of a simple AOM-based power lock for a diode laser. The ``high-frequency" power fluctuations are investigated in the frequency domain by the RPN for Fourrier frequencies between 1 Hz and 1 MHz. The power flicker noise is reduced by 32 dB until 100 Hz Fourier frequency. This noise level leads to a reduction of the RPN contribution to the clock frequency instability at 1~s averaging time at the level of 2.5$\times10^{-14}$ instead of 5$\times10^{-13}$ when laser powers are not stabilized.

The slow fluctuations are studied in the time domain by means of the Allan deviation for averaging times between 1 s and 10 000 s.  The measurement is limited by the instrumentation at short averaging times $\tau$. A minimum at the level of $5-6 \times 10^{-7}$ is obtained from $\tau=200$~s to $\tau=1\,000$~s. At longer times the instability increases, related to a power drift. A first drift was explained by temperature sensitive birefringent effects in the AOM. The second drift is  due to the temperature sensitivity of the half-wave plate combined with a PBS. This component must be replaced, as far as possible, by a polarization-and temperature insensitive beamsplitter. 
The residual drift is probably due to thermal photo-diode sensitivity. Nevertheless, the power instability reaches now the 2$\times10^{-6}$ level at $\tau=10\,000$ s, i.e. a reduction by almost three orders of magnitude by respect to the free running case.  
It is worth to note that such a power instability could still be limiting for a CPT clock based on a continuous interrogation \cite{Abdel:JAP:2015}, but not for a clock based on a Ramsey interrogation thanks to a new technique known as autobalanced Ramsey spectroscopy \cite{Sanner:2018, Yudin:2018} which can dramatically reduce laser-power effect on the frequency in CPT-Ramsey clocks \cite{Hafiz:PRAP:2018,Hafiz:APL:2018}.
Finally, we have shown that a high power stability can be reached with a simple device based on a single AOM. Such a report can be useful in many laboratories working in various fields like atomic physics, optics, sensors, or metrology, in order to know what results can be  achieved in frequency and time domains with this scheme, and what are the issues.

\section*{Acknowledgment}

We are grateful to Jos\'e Pinto Fernandes for his skill in electronics and his patience. We thank P. Yun, R. Bouchand, and Y. Le Coq for valuable contributions. We also thank V. Giordano and F. Du Burck for helpful comments. We are indebted to R. Boudot for valuable discussions and careful reading of the manuscript. F. T. was supported by French MoD, Direction G\'en\'erale de l'Armement (DGA), and THALES-AVS.

\end{document}